\documentclass[runningheads]{llncs}
\usepackage{multirow}
\usepackage{bigstrut}
\usepackage{subfigure}
\usepackage{graphicx}
\usepackage{stfloats}
\usepackage{amssymb}
\usepackage{amsopn}
\usepackage{authblk}
\usepackage{marvosym}
\usepackage{lipsum}
\usepackage{cite}
\usepackage{booktabs}  
\usepackage{threeparttable}  
\usepackage{graphicx}
\usepackage{amsmath}
\usepackage[T1]{fontenc}
\usepackage{float}
\usepackage{multirow}

\usepackage[colorlinks,linkcolor=blue,
anchorcolor=blue,
citecolor=blue,urlcolor=blue]{hyperref}
\begin{document}
	\bibliographystyle{plain}
	\title{\Large\bfseries Invariant Content Synergistic Learning for Domain Generalization of Medical Image Segmentation}
    \author{Yuxin Kang\inst{1} \and Hansheng Li\inst{1} \and Xuan Zhao\inst{1} \and  Dongqing Hu\inst{1} \and Feihong Liu\inst{1} \and Lei Cui\inst{1(\textrm{\Letter})}  \and Jun Feng\inst{1} \and Lin Yang\inst{1}}
	
	%
	\titlerunning{Invariant Content Synergistic Learning} 
	
	\authorrunning{Y. Kang et al.}
	%
	\institute{School of Information Science and Technology, Northwest University, Xi'an, 710127, Shaanxi, China}
	\maketitle
	\begin{abstract}
While achieving remarkable success for medical image segmentation, deep convolution neural networks (DCNNs) often fail to maintain their robustness when confronting test data with the novel distribution. 
To address such a drawback, the inductive bias of DCNNs is recently well-recognized.
Specifically, DCNNs exhibit an inductive bias towards image style (e.g., superficial texture) rather than invariant content (e.g., object shapes).
In this paper, we propose a method, named Invariant Content Synergistic Learning (ICSL), to improve the generalization ability of DCNNs on unseen datasets by controlling the inductive bias. First, ICSL mixes the style of training instances to perturb the training distribution. That is to say, more diverse domains or styles would be made available for training DCNNs. Based on the perturbed distribution, we carefully design a dual-branches invariant content synergistic learning strategy to prevent style-biased predictions and focus more on the invariant content. Extensive experimental results on two typical medical image segmentation tasks show that our approach performs better than state-of-the-art domain generalization methods.
	\end{abstract}
	\section{Introduction}\label{intro}
	Deep Convolution Neural Networks (DCNNs) based methods have achieved remarkable performance for medical image segmentation\cite{b1,b10,b11,b14}.
	Traditional DCNNs are trained based on the assumption that training (source) and test (target) data are identically distributed \cite{b2,b3}. 
	However, this assumption does not always hold in reality.
	In a realistic application, the medical images are in different style distribution because they are collected from different institutions where scanners and imaging protocols are different \cite{b1,b11}.
	The performance of DCNNs methods degrades significantly due to distribution gap \cite{b1,b10,b11}.
\begin{figure*}[htbp] 
	\centering	 	
	\setlength{\abovecaptionskip}{-0.05cm}
	\setlength{\belowcaptionskip}{-0.6cm} 	  
	\includegraphics[width=0.93\textwidth,height=0.20\textheight]
	{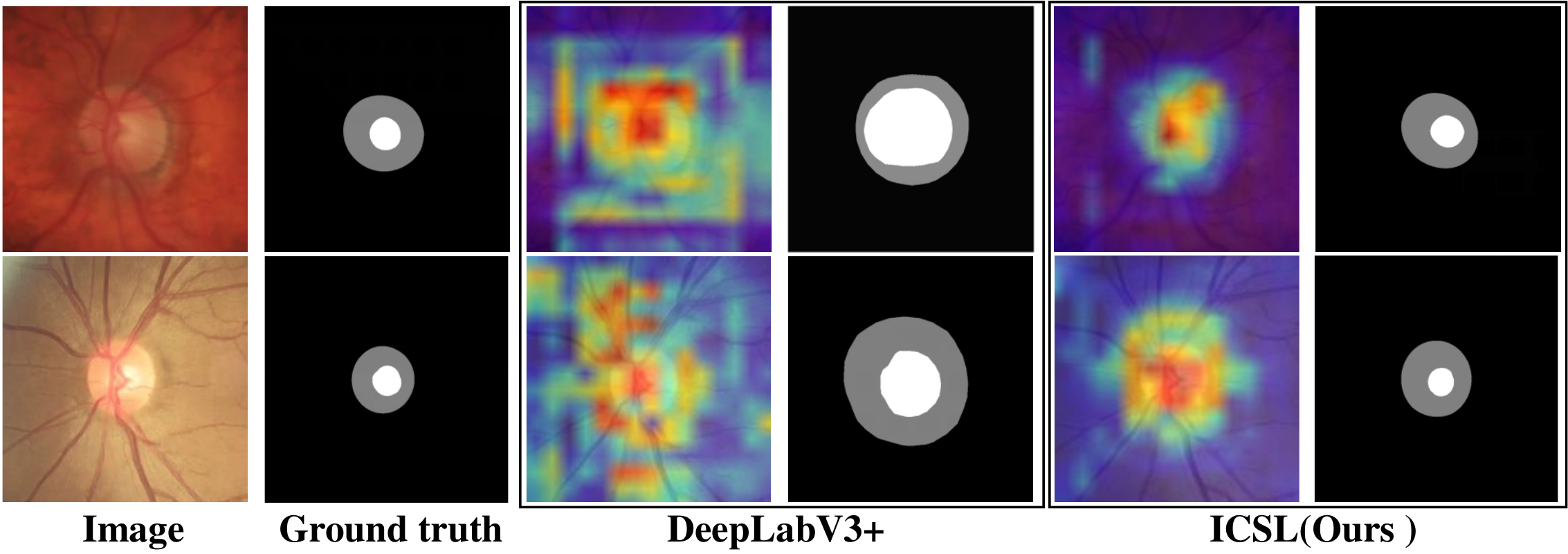} 	  	
	\caption{
		The visualization of class activation maps (CAMs) \cite{b23} and segmentation results for DeepLabV3+ \cite{b17} compared with our method on optic cup/disc (OC/OD) segmentation. DeepLabV3+ exhibits a strong bias towards image style.} 
	\label{Motivation}   
\end{figure*}
	
	To tackle this problem, researchers have explored domain adaptation (DA) \cite{b8,b9} methods to maintain the robustness of DCNNs. Despite their promising performance, DA methods have a limited clinical value due to the requirement of accessing target domain data, which is usually unavailable \cite{b10,b11}. By this, another promising direction, domain generalization (DG) methods \cite{b15,b16}, is opened up to generalize a model on unseen testing domain using several source domains. However, these methods are generally designed for classification tasks of natural image, which are difficult to be used for medical image segmentation due to the structured prediction characteristics \cite{b14}. For medical image segmentation, several latest methods explored various data augmentation techniques \cite{b13,b14} and meta-learning paradigm \cite{b10,b12} to guarantee no distribution gap between training and test datasets. Such augmentation techniques heavily depend on the quality of generated medical images. Due to the difference between generated images and real images, it will cause a dramatic performance drop if the model trained with the generated images is directly applied to real scenarios \cite{b13}. In contrast, meta-learning is highly complicated, and DCNNs would be difficult to converge when dealing with larger training datasets \cite{b15}.
	
	In stark contrast to the vulnerability of DCNNs, the human visual system generalizes well across domains.
	Especially for radiologists, they can easily recognize lesions across institutions, even medical images with the same style have not been provided to them \cite{b4}. Where does the difference come from?
	Recent studies \cite{b5,b7} has revealed that DCNNs have an inductive bias far different from human vision.
	Specifically, while radiologists tend to recognize lesions based on their invariant contents (e.g., object shapes)\cite{b7}, as shown in Fig.~\ref{Motivation}, DCNNs methods exhibit a strong bias towards image styles (e.g., color, and contrast)\cite{b5,b6}.
	This may explain why DCNNs methods are more sensitive to the distribution gap because the contrast of images is more likely to change than lesion shapes.  
	
	It is reasonable to assume a correlation between DCNNs’ inductive bias and their capability to handle distribution gaps: reducing style bias may reduces domain discrepancy.
	By this, we propose an Invariant Content Synergistic Learning (ICSL) framework which improves the domain transfer ability of DCNNs by controlling the inductive bias.
	Specifically, we first perturb the style distribution by mixing the style representation of training instances.
	Based on the perturbed distribution, we design a dual-branches invariant content synergistic learning strategy to control the inductive bias.
	In this way, DCNNs can prevent style-biased predictions and focus more on the invariant content feature. 
	We evaluate the proposed ICSL on two typical medical image segmentation tasks, i.e., optic cup/disc (OC/OD) segmentation \cite{b14} and prostate segmentation \cite{b12}. Extensive experiments validate the efficacy and universality of our approach, and ICSL outperforms several state-of-the-art DG methods.
	\begin{figure*}[tbp]  	
		\includegraphics[width=\textwidth]{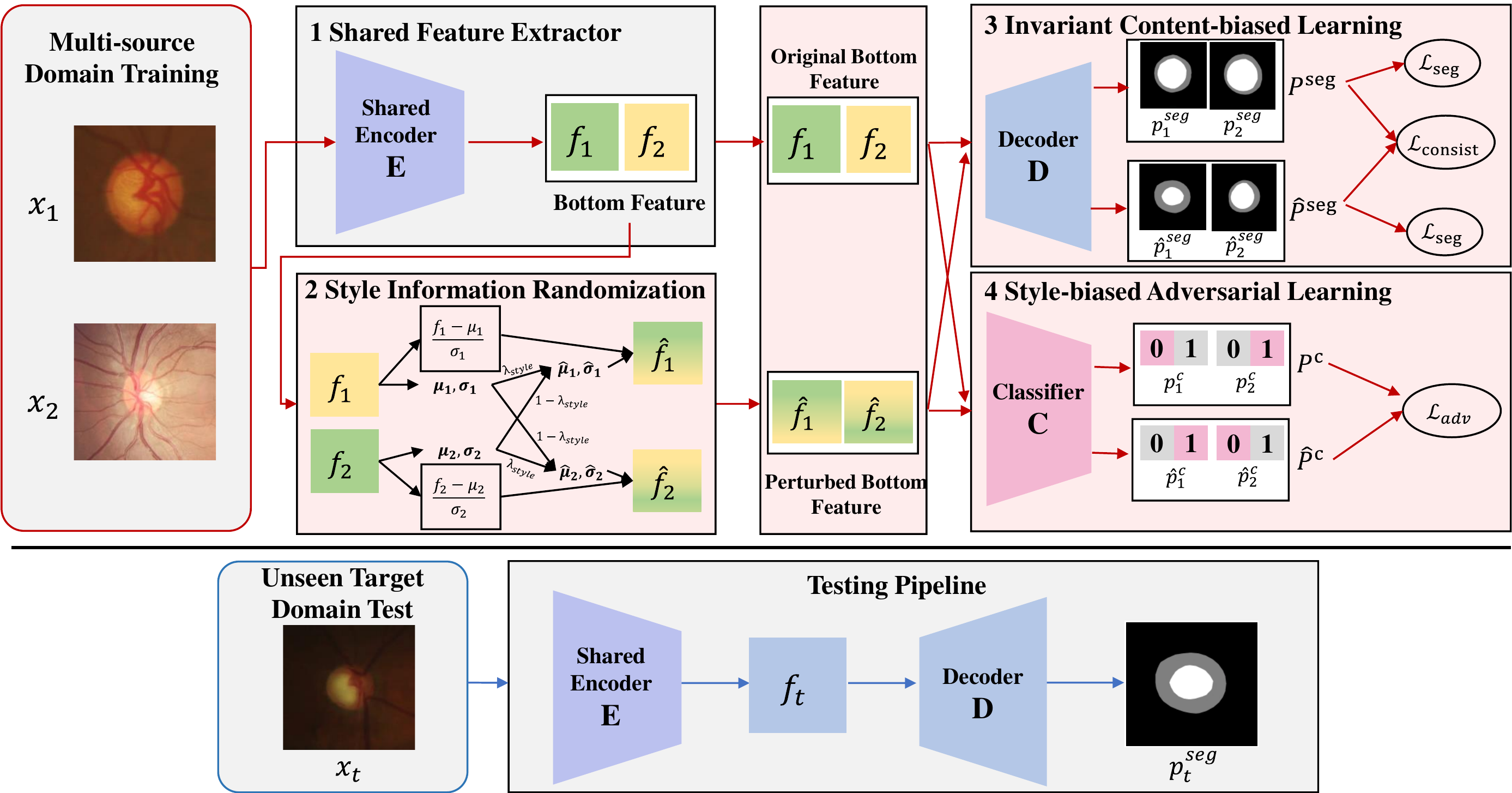} 	
		\caption{The schematic diagram of our proposed ICSL framework utilizes multi-source domain datasets for generalizable segmentation on the unseen target dataset. The red line and blue line represent the training workflow and testing workflow.} 
		\label{Network}
	\end{figure*}
	\section{Methodology}
	\subsection{Problem Definition and Method Overview} 
	\textbf{Problem Definition.}
	Let a set of $K$ source domains be denoted by $S=\left\{\left(x_{i}^{k}, y_{i}^{k} \right)_{i=1}^{N}\right\}_{k=1}^{K}$, where $x_{i}^{k}$ and $y_{i}^{k}$ donate the $i$-th image in the $k$-th source domain and the segmentation mask of $x_{i}^{k}$. The goal of our ICSL is to learn a segmentation model $f_{\theta}: x \rightarrow y$ using the $K$ distributed source domains, such that it can directly generalize to a completely unseen testing domain $T$ with a high performance.
	~\\
	\textbf{Overview.}
We present a general framework ICSL to make the DCNNs agnostic to style but focused more on content. It contains four components: a shared encoder $E$, a style information randomization (SIR) module, an invariant content-biased learning (ICL) branch, and a style-biased adversarial learning (SAL) branch. 
SIR is designed to increase the distribution diversity during training by perturbing the style information of source domain training instances.
That is, more diverse domains/styles would be made available for training a more domain-generalizable model. 
Meanwhile, there is an old saying that "two heads are better than one."
On the one hand, based on original and perturbed distribution, the ICL branch is encouraged to exploit image contents when making decisions by consistency prediction constraint(see Sect.~\ref{ICL}).
On the other hand, we also exploit the SAL branch as a discriminator, making the shared encoder $E$ less style-biased and more content-biased (see Sect.~\ref{ICL}). The pipeline of the proposed ICSL framework is illustrated in Fig.~\ref{Network}.

	\subsection{Style Information Randomization}\label{SIR}
	Our SIR is motivated by the observation that exchanging the instance-level bottom feature statistics (i.e., channel-wise mean and standard deviation) across training instances can transfer the style information while preserving the content \cite{b22}. 
	It is reasonable to assume that mixing the statistics would result in new styles information. 
	Concretely, our SIR randomly selects two instances of input training datastes, and then adopts a probabilistic convex combination between the statistics of them to generate new style information representation.
	The computations can be summarized into two steps: style normalization and style randomization.
	~\\
	\textbf{Style Normalization.} 
	Given an input training image $x_{1}$ and a randomly selected image $x_{2}$ from $K$ source domains, we first extract their bottom feature maps $f_{1}, f_{2}\in \mathbb{R}^{D\times H\times W}$ from the shared encoder $E$. $H$ and $W$ indicate spatial dimensions, and $D$ is the number of channels. Then, we compute the channel-wise mean and standard deviation $\mu_{1},\sigma_{1} \in \mathbb{R}^{D} $as style statistics \cite{b22} of $f_{1}$ by:
	\begin{align}
		\setlength{\belowdisplayskip}{-1.5cm}	
		\label{equ1}
		\mu_{1}&= \frac{1}{HW}\sum\limits_{h=1}^{H}\sum\limits_{w=1}^{W} f_{1}^{hw},\\
		\sigma_{1}&= \sqrt{\frac{1}{HW}\sum\limits_{h=1}^{H}\sum\limits_{w=1}^{W} \left ( f_{1}^{hw}-\mu_{1}  \right)^{2}+\epsilon}.
	\end{align}
    We also compute the style statistics $\mu_{2}, \sigma_{2} \in \mathbb{R}^{D}$ as style statistics of $f_{2}$.
    ~\\
	\textbf{Style Randomization.}
	Furthermore, the SIR module constructs a randomized style $\hat{\mu}_{1}, \hat{\sigma}_{1} \in \mathbb{R}^{D}$ by
	interpolating between the styles statistics of $f_{1}$ and $f_{2}$. Specifically, we convexly combine the style statistics $\mu_{1}, \sigma_{1}$ with $\mu_{2},\sigma_{2}$ and generate the mixed style statistics by:
	\begin{align}\label{equ3}
		\setlength{\belowdisplayskip}{-1cm}		
		\hat{\mu}_{1}=\lambda_{style}\cdot\mu_{1}+(1-\lambda_{style})\cdot\mu_{2},\\
		\hat{\sigma}_{1}=\lambda_{style}\cdot\sigma_{1}+(1-\lambda_{style})\cdot\sigma_{2},
	\end{align}
	where $\lambda_{style} \sim Uniform(0, 1)$ is a random interpolation weight. Finally, the mixed statistics are applied to $f_{1}$ by:
	\begin{align}\label{equ4}	
		\hat{f}_{1}=\hat{\sigma}_{1}\cdot\frac{f_{1}-\mu_{1}}{\sigma_{1}}+\hat{\mu}_{1}.
	\end{align}
    In this way, for each feature map $f_{n}$ from $n$-th input training sample, a perturbed feature $\hat{f}_{n}$ will be generated during training. 
    \vspace{-0.3cm}
	\subsection{Dual-branches Invariant Content Synergistic Learning}
	\textbf{Invariant Content-biased Learning.}\label{ICL} 
	Based on the perturbed training distribution, we further design invariant content-biased learning branch, which no longer relies on the style but focus more on the content in making a decision.
	The critical point of this branch is the consistency prediction assumption. Specifically, the segmentation results should remain the same (i.e., based on invariant content), no matter how the style changed.
	As stated above, we design a shared decoder $D$ to obtain the segmentation results of original and perturbed features.
	On the one hand, we encourage the shared encoder $E$ and decoder $D$ to segment each feature rightly by minimizing a supervised segmentation loss $\mathcal L_{seg}$.
	On the other hand, we design a consistency loss $\mathcal L_{consist}$ to minimize the discrepancy among the predictions at original and perturbed feature.
	\begin{align}\label{equ5}
		\setlength{\belowdisplayskip}{-1cm}	
		\underset{E,D}{min} \mathcal L_{seg}= \frac{1}{N}\sum\limits_{n=1}^{N}&\frac{\mathcal L_{ce}(p^{seg}_{n},y_{n})+\mathcal L_{ce}(\hat{p}^{seg}_{n},y_{n})}{2},\\
		\underset{E,D}{min} \mathcal L_{consist}= &\frac{1}{N}\sum\limits_{n=1}^{N}||p^{seg}_{n}-\hat{p}^{seg}_{n}||_{2},
	\end{align}
	where $p^{seg}_{n}$ and $\hat{p}^{seg}_{n}$ are the segmentation result of $f_{n}$ and $\hat{f}_{n}$. $N, y_{n}$ donate the input mini-batch size and the segmentation mask of $n$-th input sample. $\mathcal L_{ce}$ and $||...||_{2} $ represent the pixel-wise cross-entropy loss and $\mathcal L_{2}$ loss \cite{b14}.
	The $\mathcal L_{2}$ loss requires each pixel value in $p^{seg}_{n}$ to be consistent with the corresponding value of the
	same pixel in $\hat{p}^{seg}_{n}$, making the performance of model less vulnerable to the varying style representation \cite{b14}.
	~\\
	\textbf{Style-biased Adversarial Learning.}\label{SAL}  
	In addition, we constrain the shared encoder $E$ from learning style-biased representation by adopting an adversarial learning branch. 
	In other words, we make the encoder $E$ incapable of discriminating the style of input training instances.
	To achieve this, we build an auxiliary style classifier $C$ as a discriminator to obtain a style-biased prediction. Precisely, the classifier $C$ consists of a global average pooling layer, a Batch Normalization layer, a ReLU activation layer, a convolutional layer, and a Softmax activation layer.
	In terms of implementation, $f_{n}$ and $\hat{f}_{n}$ are taken as inputs to the classifier $C$. $C$ is trained to map the input feature into a binary style label $p^{c}_{n}$ and $\hat{p}^{c}_{n}$, where the true denotes the original style representation and false denotes the perturbed representation. The shared encoder $E$ is then trained to fool classifier $C$ so that it cannot distinguish between the two representations by minimizing an adversarial loss $\mathcal L_{adv}$ \cite{b8}:
	\begin{align}
		\setlength{\belowdisplayskip}{-1cm}	
		\label{equ7}
		\underset{E,C}{min} \mathcal L_{adv}= -\frac{1}{N}\sum\limits_{n=1}^{N} log(p^{c}_{n})-\frac{1}{N}\sum\limits_{n=1}^{N} log(1-\hat{p}^{c}_{n}).
	\end{align}
	~\\
	\textbf{Training strategy.}
	We employ DeepLabV3+ \cite{b17} with the MobileNetV2 \cite{b18} backbone as the basic segmentation framework.
	The training images are randomly chosen from the multiple source domains to form a training batch.
	The overall loss function to train ICSL is:
	\vspace{-0.2cm}
	\begin{align}
		\setlength{\belowdisplayskip}{-1cm}	
		\label{equ9}
		\underset{E,C,D}{min} \mathcal L_{all}= \mathcal L_{seg}+\mathcal L_{consist}+\lambda_{adv}\mathcal L_{adv},
	\end{align}
	where $\lambda_{adv}$ is a weight coefficient. We first trained a vanilla segmentation network without $\mathcal L_{consist}$ and $\mathcal L_{adv}$ for 40 epochs with a learning rate of 1e$-$3.
	Then the whole framework is trained for another 80 epochs with an initial learning rate of 1e$-$3 and a batch size of 16. 
	The loss function weight $\lambda_{adv}$ was set as 0.2.
	Our framework was built on PyTorch and trained on one NVIDIA RTX TITAN GPU. We adopted Adam optimizer to train the model. 
	The final prediction at test time is made by the shared encoder $E$ followed with decoder $D$.
\begin{table*}[tbp]
	\setlength{\belowcaptionskip}{-0.cm} 
	\centering
	\caption{Comparison with recent DG methods on the OC/OD segmentation task. The top two values are emphasized using bold and underline, respectively.}
	\label{table2}%
	\LARGE  
	\resizebox{\linewidth}{!}{
		\begin{tabular}{c|ccccc|ccccc|c|ccccc|ccccc|c}
			\hline
			\hline
			\textbf{Task} & \multicolumn{5}{c|}{\textbf{Optic Cup Segmentation}} & \multicolumn{5}{c|}{\textbf{Optic Disc Segmentation}} & \multirow{2}[4]{*}{\textbf{Overall}} & \multicolumn{5}{c|}{\textbf{Optic Cup Segmentation}} & \multicolumn{5}{c|}{\textbf{Optic Disc Segmentation}} & \multirow{2}[4]{*}{\textbf{Overall}} \\
			\cline{1-11}\cline{13-22}    \textbf{Domain}  & 1  & 2  & 3  & 4  & Avg.  & 1  & 2  & 3  & 4  & Avg.  &       & 1  & 2  & 3  & 4  & Avg.  & 1  & 2  & 3  & 4  & Avg.  &  \\
			\hline
			\hline
			& \multicolumn{11}{c|}{\textbf{Dice Coefficient (Dice) $\uparrow$} }                                      & \multicolumn{11}{c}{\textbf{Average Surface Distance (ASD) $\downarrow$}} \\
			\hline
			\hline
			Baseline\cite{b17}& 77.03  & 78.21  & 80.28  & 84.74  & 80.07  & 94.96  & 89.69  & 89.33  & 90.09  & 91.02  & 85.54  & 21.64  & 16.77  & 11.58  & 7.92  & 14.48  & 8.98  & 17.10  & 12.64  & 9.10  & 11.96  & 13.22  \\
			CutMix(2019)\cite{b19} & 76.97  & 81.02 & 83.42  & 86.83  & 82.06  & 93.83  & 91.97  & 90.13  & 88.79  & 91.18  & 86.62  & 22.38  & 12.65  & 13.33  & 7.94  & 14.08  & 9.40  & 12.83  & 14.41  & 11.80  & 12.11  & 13.09  \\
			SAML(2020)\cite{b10} & 83.16  & 75.68  & 82.00  & 82.88  & 80.93  & 94.30  & 91.17  & 92.28  & 87.95  & 91.43  & 86.18  & 18.20  & 16.87  & 13.38  & 9.89  & 14.59  & 10.08  & 12.90  & 12.31  & 14.22  & 12.38  & 13.48  \\
			DoFE(2020)\cite{b14}& 83.59  & 80.00  & \underline{86.66}  & 87.04  & 84.32  & \textbf{95.59} & 89.37  & 91.98  & 93.32  & 92.57  & 88.44  & 17.68  & 14.71  & \underline{9.22}  & 7.21  & 12.38  & \textbf{7.23} & 14.08  & 11.43  & 9.29  & 10.51  & 11.44  \\
			FedDG(2021)\cite{b12}& 84.13  & 71.88  & 83.94  & 85.51  & 81.37  & 95.37  & 87.52  & 93.37  & \underline{94.50}  & 92.69  & 87.03  & -     & -     & -     & -     & -     & - & -     & -     & -     & -     & - \\
			FACT(2021) \cite{b16}& 79.66  & 79.25  & 83.07  & 86.57  & 82.14  & 95.28  & 90.19  & \textbf{94.09 } & 90.54  & 92.53  & 87.33  & 20.71  & 14.51  & 11.61  & 7.83  & 13.67  & 8.19  & 14.71  & \textbf{8.43} & 10.57  & 10.48  & 12.07  \\
			\hline
			w/o SIR & 83.57 & 80.98 & 85.75 & 86.12 & 84.06 & 93.98  & 91.72 & 92.48  & 93.24 & 92.86 & 88.46 & 16.48 & 11.96 & 10.21 & 7.83  & 11.62 & 8.04  & 12.37 & 9.84  & 7.91 & 9.54 & 10.58 \\
			w/o ICL & \underline{84.27} & 81.33 & 86.40 & 86.78 & \underline{84.70} & 94.77  & 91.36 & 93.76  & 94.12 & 93.50 & 89.10 & 15.93 & \textbf{10.37} & 9.84 & \underline{7.00}  & 10.79 & \underline{7.25}  & 12.66 & 8.76  & 7.52 & 9.05 & 9.92 \\
			w/o SAL & 83.42 & \underline{81.60} &  86.15 & \underline{87.11} & 84.57 & 94.91  & \underline{93.36} & 92.94  & 93.48 & \underline{93.67} & \underline{89.12} & \underline{14.10} & \underline{10.54} & 9.62 & 7.39  & \underline{10.41} & 7.28  & \underline{9.52} & 9.38  & \textbf{5.37} & \underline{7.89} & \underline{9.15} \\
			\textbf{ICSL(Ours))} & \textbf{86.23} & \textbf{81.95} & \textbf{87.47} & \textbf{87.14} & \textbf{85.70} & \underline{95.39}  & \textbf{93.49} & \underline{93.94}  & \textbf{94.88} & \textbf{94.43} & \textbf{90.06} & \textbf{14.03} & 11.44 & \textbf{9.21} & \textbf{6.73}  & \textbf{10.36} & 7.88  & \textbf{9.36} & \underline{8.60}  & \underline{5.68} & \textbf{7.88} & \textbf{9.11} \\
			\hline
			\hline
	\end{tabular}}%
\end{table*}%
	\vspace{-0.4cm} 
	\section{Experiments} 
	\vspace{-0.3cm} 
	\textbf{Datasets and Evaluation Metrics.}
	We evaluate our method on two medical image segmentation tasks, i.e., OC/OD segmentation on retinal fundus images and prostate segmentation on T2-weighted MRI. For OC/OD segmentation, the public dataset contains 1070 cases from
	4 clinical centers \cite{b14}. For prostate segmentation, the public dataset contains 116 T2-weighted MRI cases from
	6 domains \cite{b12}. The statistics of two datasets are summarized in Table.1 of supplementary material. For pre-processing of both datasets, we resize the image to 384 $\times$ 384 as network input. We randomly partition each domain into training and testing sets based on a ratio of 4: 1. Data augmentation of random rotation, scaling, and flipping are employed in the two tasks. For evaluation, we adopt two commonly-used metrics, including the Dice coefficient (Dice) and Average Surface Distance (ASD) \cite{b14} for boundary agreement assessment. The higher Dice and lower ASD indicate a better performance.
	~\\
	\textbf{Comparison with DG methods.}
	Our experiments follow the practice in DG literature to adopt the leave-one-domain-out strategy, i.e., training on $K-1$ distributed source domains and testing on the one left-out unseen target domain. 
	For the baseline, we treated all the source domains as one domain and trained the segmentation network DeepLabV3+ in a usual supervised way. 
	We first compare our method with recent network regularization methods, such as CutMix \cite{b19} and FACT \cite{b16}. We also compare with state-of-the-art DG methods for OC/OD and Prostate MRI segmentation, i.e.,  DoFE \cite{b14},  FedDG \cite{b12}, SAML \cite{b16}.
	\begin{table}[tbp]
		\centering
		\caption{Comparison with recent DG methods on Prostate MRI segmentation.}
		\label{table3}%
		\resizebox{\textwidth}{!}{
			\begin{tabular}{c|ccccccc|ccccccc}
				\hline
				\hline
				\textbf{Domain}  & \textbf{1}     & \textbf{2}     & \textbf{3}     & \textbf{4}     & \textbf{5}     & \textbf{6}     & \multicolumn{1}{c|}{\textbf{Avg.}} & \textbf{1}     & \textbf{2}     & \textbf{3}     & \textbf{4}     & \textbf{5}     & \textbf{6}     & \multicolumn{1}{p{2.25em}}{\textbf{Avg.}} \\
				\hline
				\hline
				\multicolumn{1}{c|}{} & \multicolumn{7}{c|}{\textbf{Dice Coefficient (Dice) $\uparrow$ }} & \multicolumn{7}{c}{\textbf{Average Surface Distance (ASD) $\downarrow$}} \\
				\hline
				\hline
				Baseline\cite{b17} & 83.51  & 81.94  & 82.29  & 85.44  & 84.66  & 84.89  & 83.79  & 6.15  & 5.95  & 6.24  & 4.24  & 7.49  & 4.42  & 5.75  \\
				CutMix(2019) \cite{b19}& 83.27  & 84.36  & \underline{85.35}  & 81.06  & 85.67  & 84.95  & 84.11  & 5.63  & 5.71  & 5.62  & 5.75  & 6.65  & 4.27  & 5.61  \\
				SAML(2020) \cite{b10}& 89.66  & \underline{87.53}  & 84.43  & 88.67  & \textbf{87.37}  & 88.34  & \underline{87.67}  & \underline{4.11}  & \underline{4.74}  & 5.40  & \underline{3.45}  & \underline{4.36}  & \underline{3.20}  & \underline{4.21}  \\
				DoFE(2020) \cite{b14}& 84.66  & 84.42  & 85.22  & 86.31  & \underline{87.60}  & 86.96  & 85.86  & 4.95  & 5.23  & \underline{5.04}  & 4.30  & \textbf{4.23}  & 3.33  & 4.51  \\
				FedDG(2021) \cite{b12}& \underline{90.19}  & 87.17  & 85.26  & 88.23  & 83.02  & \textbf{90.47} & 87.39  & - & - & - & - & - & - & - \\
				FACT(2021) \cite{b16}& 85.82  & 85.45  & 84.93  & \underline{88.75}  & 87.31  & 87.43  & 86.62  & 4.77  & 5.60  & 6.10  & 3.90  & 4.37  & 3.43  & 4.70  \\
				\hline
				\textbf{ICSL(Ours)} & \textbf{91.26} & \textbf{90.22} & \textbf{87.60} & \textbf{88.88} & 85.15  & \underline{89.52}  & \textbf{88.77} & \textbf{3.10} & \textbf{3.76} & \textbf{4.16} & \textbf{3.01} & 4.43  & \textbf{2.99} & \textbf{3.57} \\
				\hline
				\hline
		\end{tabular}}%
	\end{table}%
	
	Table.~\ref{table2} shows the quantitative results on the OC/OD segmentation tasks.
	We see that different DG methods can improve the overall generalization performance more or less over baseline.
	It attributes to their augmentation effect or meta-learning to extract general representation.
	Compared with these methods, our ICSL achieves higher overall performance and improves Dice and ASD for both optic disc and cup segmentation. In particular, our approach largely surpasses the strongest competitor, DoFE, by a large margin (1.62\% average Dice and 2.33 average ASD). This benefits from our ICSL, which increases the style diversity during training. Specifically, for other DG methods, their learning can only access the training distribution and fail to extract features towards invariant content in a diverse distribution. In contrast, our method enables DCNNs to take full advantage of the multi-source distributions and prevent style-biased predictions. It is also worth noting that while all the compared methods except CutMix exploit additional layers on the baseline at test time, ICSL does not require any extra parameters or computations over the baseline.

\begin{figure}[tbp] 
		\centering	
		\setlength{\abovecaptionskip}{-0.1cm}
		\setlength{\belowcaptionskip}{-0.4cm} 
		\includegraphics[width=0.92\textwidth]{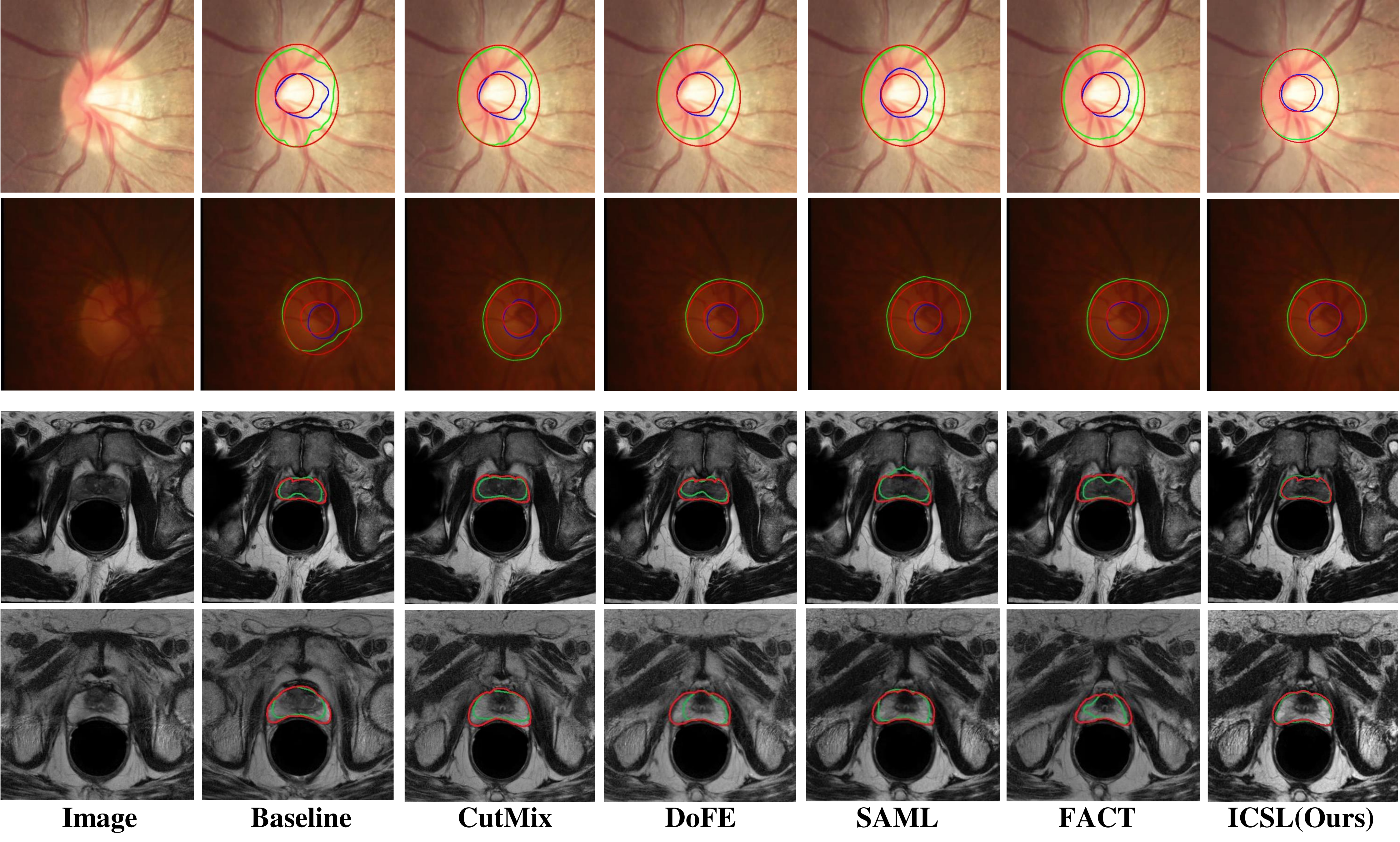} 	
		\caption{Qualitative comparison in OC/OD segmentation (top two rows) and prostate segmentation (bottom two rows). The green and blue contours indicate the prediction of OD, OC and prostate, respectively. All red contours represent the ground truths.} 
		\label{SHOW2} 	
	\end{figure}

\begin{figure}[tbp] 
		\centering	
		\setlength{\abovecaptionskip}{-0.1cm}
		\setlength{\belowcaptionskip}{-0.6cm} 
		\includegraphics[width=0.8\textwidth,height=0.20\textheight]{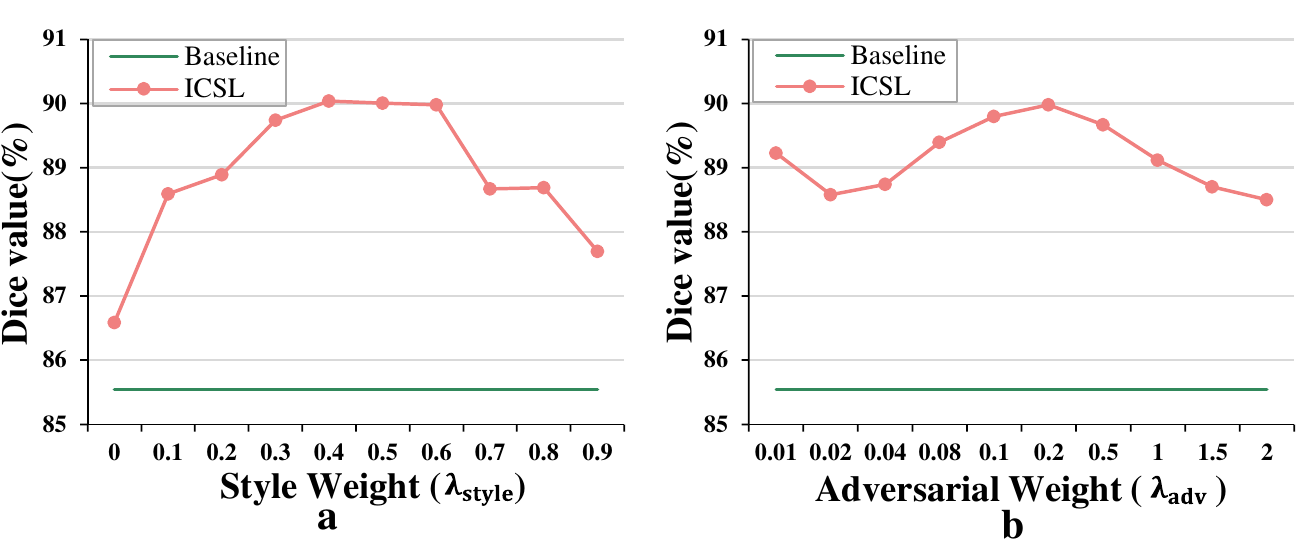} 	
		\caption{Average Dice of ICSL on fundus image segmentation with varying (a) style weight $\lambda_{style}$ and (b)  adversarial weight $\lambda_{adv}$.} 
		\label{SHOW8} 	
	\end{figure}

	For prostate MRI segmentation, as shown in Table.~\ref{table3}, the DG methods generally perform better than baseline, and the improvements are relatively significant. It may be because the difficulty is mitigated among gray scale images. Our ICSL obtains the highest average Dice and ASD across the six unseen sites. Specifically, our approach outperforms the strongest competitor SAML by a large margin (1.1\% average Dice and 0.64 average ASD).
	~\\
	\textbf{Qualitative Results.}
	We present two sample segmentation results from each task in Fig.~\ref{SHOW2} for better visualization. As we can see, by reducing the style inductive bias, the predictions of our approach have smoother contours and are closer to the ground truth than others. Specifically, in the second row, it is pretty hard to distinguish OD and OC for other methods due to the low image contrast, while our method can still segment OC and OD with accurate boundaries.
	~\\
	\textbf{Ablation Study and Discussion.} 
	We first validate the effect of three critical components in our method by removing them (i.e., w/o SIR, ICL, or SAL) and observing the model performance. As shown in Table.~\ref{table2}, removing SIR (i.e., $\lambda_{style}= 0$) will lead to a decrease in the generalization performance. The reason is that removing SIR will not reinforce the ability of DCNNs to cope with unseen styles.
	Furthermore, experiments without ICL and SAL verify the effectiveness of each component of ICSL. 
	We also analyze the effect of style weight $\lambda_{style}$ and adversarial coefficient $\lambda_{adv}$ on our model performance. As shown in Fig.~\ref{SHOW8}(a) and (b), setting $\lambda_{style} \textless 1$ or $\lambda_{adv} \textgreater 0$ can always improve the model performance (pink line). The actual performance peaks around $\lambda_{style} = 0.4$ or $\lambda_{adv} = 0.2$. The result shows that mixing (i.e., $\lambda_{style} \textgreater 0$) is better than exchanging (i.e., $\lambda_{style} = 0$). Meanwhile, we suggest that weak adversarial learning (i.e., $\lambda_{adv} \textless 0.2$) maybe less helpful in reducing style bias, strong adversarial learning (i.e., $\lambda_{adv} \textgreater 0.2$) may complicate the optimization and damage the performance.
	\vspace{-0.7cm} 
	\section{\large\bfseries{Conclusion}}
	\vspace{-0.25cm}
	We present an Invariant Content Synergistic Learning (ICSL) framework robust against the distribution gap caused by style variability across domains. By controlling the inductive bias, ICSL is trained to rely more on content rather than style in the decision-making process. Extensive experiments results support our hypothesis and demonstrate the effectiveness of our proposed method. The principle of how we deal with the inductive biases of DCNNs would be further applied to other areas, such as defending against adversarial attacks \cite{b24}.
	

\begin{thebibliography}{22}
		\bibitem{b1}L. Zhang, X. Wang, D. Yang, T. Sanford, S. Harmon, B. Turkbey, B.J. Wood, H. Roth, A. Myronenko, D. Xu, and Z. Xu, "Generalizing Deep Learning for Medical Image Segmentation to Unseen Domains via Deep Stacked Transformation," in IEEE Transactions on Medical Imaging, vol. 39, no. 7, pp. 2531-2540, July 2020, doi: 10.1109/TMI.2020.2973595.
		
		\bibitem{b2}D. S. W. Ting, L. R. Pasquale, L. Peng, J. P. Campbell, A. Y. Lee, R. Raman, G. S. W. Tan, L. Schmetterer, P. A. Keane, and T. Y. Wong, "Artificial intelligence and deep learning in ophthalmology," in British Journal of Ophthalmology, vol. 103, no. 2, pp. 167–175, 2019.
		
		\bibitem{b3}S. Wang, L. Yu, X. Yang, C.-W. Fu, and P.-A. Heng, "Patch-based output space adversarial learning for joint optic disc and cup segmentation," in IEEE Transactions on Medical Imaging, vol. 38, no. 11, pp. 2485–2495, 2019.
		
		\bibitem{b4}P. A. Ganea, M. B. Pickard, and J. S. DeLoache, "Transfer between picture books and the real world by very young children," in Journal of Cognition and Development, vol. 9, no. 1, pp. 46-66, 2008, doi: 10.1080/15248370701836592.
		
		\bibitem{b5}N. Baker, H. Lu, G. Erlikhman, and P. J. Kellman, "Deep convolutional networks do not classify based on global object shape," in PLOS Computational Biology, 2018.
		
		\bibitem{b6}R. Geirhos, P. Rubisch, C. Michaelis, M. Bethge, F. A. Wichmann, and W. Brendel, "Imagenet-trained cnns are biased towards texture; increasing shape bias improves accuracy and robustness," in International Conference on Learning Representations (ICLR), 2019.
		
		\bibitem{b7}K. Hermann, T. Chen, and S. Kornblith, "The origins and prevalence of texture bias in convolutional neural networks," in Advances in Neural Information Processing Systems, pp. 19000--19015, 2020.
		
		\bibitem{b8}Q. Dou, C. Ouyang, C. Chen, H. Chen, and P.-A. Heng, "Unsupervised cross-modality domain adaptation of ConvNets for biomedical image segmentations with adversarial loss," in International Joint Conference on
		Artificial Intelligence (IJCAI), 2018.
		
		\bibitem{b9}Y. Zhang, S. Miao, T. Mansi, and R. Liao, "Task driven generative modeling for unsupervised domain adaptation: Application to X-ray image segmentation," in International Conference on Medical Image Computing and Computer-Assisted Intervention (MICCAI), Springer, pp. 599–607, 2018.
		
		\bibitem{b10}Q. Liu, Q. Dou, and P.-A. Heng,  "Shape-aware meta learning for generalizing prostate MRI segmentation to unseen domains," in International Conference on Medical Image Computing and Computer Assisted Intervention (MICCAI), Springer, 2020, pp. 475–485.
		
		\bibitem{b11}X. Liu, S. Thermos, A. O’Neil, and S. Tsaftaris, "Semi-supervised meta-learning with disentanglement for domain-generalised medical image segmentation," in International Conference on Medical Image Computing and Computer Assisted Intervention (MICCAI),  Springer, 2021, pp. 307--317.
		
		\bibitem{b12}Q. Liu, C. Chen, J. Qin, Q. Dou and P. -A. Heng, "FedDG: Federated domain generalization on medical image segmentation via episodic learning in continuous frequency space," in 2021 IEEE/CVF Conference on Computer Vision and Pattern Recognition (CVPR), 2021, pp. 1013-1023, doi: 10.1109/CVPR46437.2021.00107. 
		
		\bibitem{b13}F. Lv, T. Liang, X. Chen and G. Lin, "Cross-domain semantic segmentation via domain-invariant interactive relation transfer," in 2020 IEEE/CVF Conference on Computer Vision and Pattern Recognition (CVPR), 2020, pp. 4333-4342, doi: 10.1109/CVPR42600.2020.00439. 
		
		\bibitem{b14}S. Wang, L. Yu, K. Li, X. Yang, C. -W. Fu and P. -A. Heng, "DoFE: domain-oriented feature embedding for generalizable fundus image segmentation on unseen datasets," in IEEE Transactions on Medical Imaging, vol. 39, no. 12, pp. 4237-4248, Dec. 2020, doi: 10.1109/TMI.2020.3015224. 
		
		\bibitem{b15}K. Zhou, Y. Yang, T. Hospedales, and T. Xiang, "Deep domain-adversarial image generation for domain generalization," in 2020 Association for the Advancement of Artificial Intelligence (AAAI), pp. 13025–13032, 2020.
		
		\bibitem{b16}Q. Xu, R. Zhang, Y. Zhang, Y. Wang and Q. Tian, "A fourier-based framework for domain generalization," 2021 IEEE/CVF Conference on Computer Vision and Pattern Recognition (CVPR), 2021, pp. 14378-14387, doi: 10.1109/CVPR46437.2021.01415. 
		
		\bibitem{b17}L.-C. Chen, Y. Zhu, G. Papandreou, F. Schroff, and H. Adam, "Encoder decoder with atrous separable convolution for semantic image segmentation," in Proceedings of the European conference on computer vision (ECCV), 2018, pp. 801–818.
		
		\bibitem{b18}M. Sandler, A. Howard, M. Zhu, A. Zhmoginov, and L.-C. Chen, "MobileNetV2: Inverted residuals and linear bottlenecks," in Proceedings of the IEEE Conference on Computer Vision and Pattern Recognition, 2018, pp. 4510–4520.
		
		
		\bibitem{b19}S. Yun, D. Han, S. J. Oh, S. Chun, J. Choe, and Y. Yoo, "CutMix: Regularization strategy to train strong classifiers with localizable features," in International Conference on Computer Vision (ICCV), 2019. 
		
		\bibitem{b20}F. M. Carlucci, A. D’Innocente, S. Bucci, B. Caputo, and T. Tommasi, "Domain generalization by solving jigsaw puzzles," in Proceedings of the IEEE Conference on Computer Vision and Pattern Recognition, 2019, pp. 2229–2238. 
		
		\bibitem{b21}L. v. d. Maaten and G. Hinton, "Visualizing data using t-SNE," Journal of machine learning research, vol. 9, no. Nov, pp. 2579–2605, 2008.
		
		\bibitem{b22}X. Huang and S. Belongie, "Arbitrary style tansfer in real-time with adaptive instance normalization," in 2017 IEEE International Conference on Computer Vision (ICCV), 2017, pp. 1510-1519, doi: 10.1109/ICCV.2017.167. 
		
		\bibitem{b23}B. Zhou, A. Khosla, A. Lapedriza, A. Oliva and A. Torralba, "Learning deep features for discriminative localization," in 2016 IEEE Conference on Computer Vision and Pattern Recognition (CVPR), 2016, pp. 2921-2929, doi: 10.1109/CVPR.2016.319.
		
		\bibitem{b24}I. J. Goodfellow, J. Shlens, and C. Szegedy, "Explaining and harnessing adversarial examples," in 2015 International Conference on Learning Representations (ICLR), 2015.
		
	\end{thebibliography}
\end{document}